\documentclass{article}
\usepackage{srcltx}
\usepackage[english]{babel}
\usepackage{amsmath}
\usepackage{amssymb}
\usepackage{amsfonts}
\usepackage{latexsym}
\usepackage{textcomp}
\usepackage{appendix}
\usepackage{multirow}
\usepackage{booktabs}
\usepackage{rotating}
\usepackage{afterpage}
\usepackage{pdflscape}
\usepackage{url}
\usepackage{array}
\usepackage{marvosym}
\usepackage{graphics}
\usepackage{graphicx}
\usepackage[authoryear]{natbib}
\usepackage[table]{xcolor}
\usepackage[margin=1.75in]{geometry}
\usepackage{epsfig}
\usepackage{float}
\usepackage{authblk}

\usepackage[caption=false]{subfig}

\title{Time-frequency co-movements between commodities and economic policy uncertainty across different crises}

\author[1]{M. Bel\'en Arouxet}
\author[2]{Aurelio F. Bariviera \thanks{aurelio.fernandez@urv.cat}}
\author[3,4]{Ver\'onica E. Pastor}
\author[4]{Victoria Vampa}
\affil[1]{\scriptsize Universidad Nacional de La Plata, Facultad de Ciencias Exactas,\\ Centro de Matemática de La Plata, Argentina}
\affil[2]{\scriptsize  Universitat Rovira i Virgili, Department of Business, ECO-SOS,\\ Av. Universitat 1, 43204 Reus, Spain}
\affil[3]{\scriptsize  Universidad de Buenos Aires, Facultad de Ingenier\'ia, Departamento de Matem\'aticas}
\affil[4]{\scriptsize Universidad Nacional de La Plata, Facultad de Ingenier\'ia,\\ Departamento de Ciencias B\'asicas, Argentina}

\begin{document}

\maketitle

\begin{abstract}
Commodity futures constitute an attractive asset class for portfolio managers. Propelled by their low correlation with other assets, commodities begin gaining popularity among investors, as they allow to capture diversification benefits. After more than two decades of active investing experience, this paper examines the time and frequency spillovers between the Economic Policy Uncertainty \citep{Davis2016} and a broad set of commodities. The period under examination goes from December 1997 until April 2022, covering political, economic, and even health crises. We apply a wavelet coherence analysis between time series, in order to shed light on the time-frequency comovements and lead-lag relationships.
This research finds a distinct impact on the commodities, depending on the nature of the crisis. In particular, during the global financial crisis and the Covid-19 crisis, comovements are stronger in most  commodities. 
. 

\end{abstract}

\section{Introduction\label{sec:intro}}
Commodities are goods that are commonly utilized as raw materials in the creation of other goods and services. Supply and demand interactions in the globalized world play a big role in commodity pricing. Weather, geopolitical events, and supply-side shocks can all have an impact on supply and demand situations (e.g., wars and hurricanes).

In the last two decades, commodities have been gaining weight in international portfolios, due to their low or negative correlations with returns on other main asset classes. In fact, \cite{Adams2015} argues that commodities evolved into a new investment style for institutional investors. However, larger participation in portfolios induces greater market integration and information spillovers. Thus, this shift could make diversification benefits to diminish. This situation makes commodities more sensible to general macroeconomic situations. As a consequence, commodities responses to general economic situations are not only of interest for traders and investors but also for governments. For example, countries like Chile heavily depend on the mining industry. Copper represents 10.9\% of this country's Gross Domestic Product (GPD) and generates around 7.8\% of the total tax revenue \citep{CopperChile}. A stronger linkage between copper and economic uncertainty could impact the whole economy by diminishing government revenues. 


The emergence of commodities as a new financial asset introduces a new style of index investing, which could affect futures prices. The literature on this subject draws different conclusions. For example, \cite{Stoll2010} do not find causation of commodity index investing in futures prices, while \cite{IrwinSanders} report no effect on volatility.
Comovements between different commodities and stocks (regarding return or volatility) have been studied previously. For example, \cite{Nguyen2021} finds that the comovement between stocks, commodities and clean energy is comparatively high, but green bonds have a lower correlation with the other financial assets. More recently, \cite{FernandesAraujoSilvaTabak2022} study the randomness of commodities prices using information theory quantifiers. 

Additionally, some papers studied the macroeconomic determinants of commodities returns or volatilities. In this sense, \cite{Batten2010} report that gold volatility could be partially explained by monetary variables. However, other precious metals do not share this common factor, in line with \cite{Erb2006}, who argue that precious metals constitute a
heterogeneous group that cannot be considered a single asset class. \cite{Chen2015} finds a significant influence of oil price shocks on the behavior of commodity sectors in China and that domestic macro fluctuations impact  the comovements depending on the length of the holding period. Therefore, there is a necessity to study the comovements of a broader range of commodities with respect to a general index of economic uncertainty. The rationale is that, on the one hand, commodity futures represent in some way opinions for future demand of real assets, but at the same time they represent a speculative asset. 

The aim of this paper is to gain insight into the return co-movements of a wide set of commodities with the economic and Economic Policy Uncertainty measured according to the index developed by \cite{Davis2016}. This information is relevant for practitioners and policymakers. Proper knowledge of the return dynamics during distressed times could help to detect changes in portfolio diversification opportunities as well as portfolio optimization. Additionally, regulatory bodies could adapt their prudential regulation framework to mitigate contagion risks in the wake of a distressing event. 

From a methodological point of view, this work illustrates the advantages of using wavelet tools for the analysis of individual time series in time and frequency to find common powers and information about the phase relationship between two time series. Specifically, the data analyzed in this paper highlight the potential of wavelet tools to interpret time series quantitatively and better understand the oscillations of displacements in frequency space with other variables to find physical phase relationships. In summary, the application of a wavelet-based methodology to time series is useful to understand the dynamic relationship between time series.

This paper contributes to the literature in four main aspects: (i) a wide sample of commodities time series during a long period of time (1997-2022) are studied, becoming the analysis different from previous studies; (ii) it scrutinizes the dynamic interaction between Economic Policy Uncertainty (proxied by GEPU index) and agricultural, energy and metal commodities; (iii) the paper employs a wavelet-based analysis which is able simultaneously to study the interaction of the time series in both time and frequency; and (iv) it uncovers that different crises affect unevenly the commodity market. 

The remaining of the paper is organized as follows: Section \ref{sec:methods} detail fully explains the methodology; Section \ref{sec:DataResults} describes data and presents the results and their discussion. Finally, Section \ref{sec:conclusions} draws the main conclusions and presents prospective research lines.


\section{Methods \label{sec:methods}}

The continuous wavelet transform (CWT) allows us to analyze the time series in the time-frequency domain, and thus identifies localized periodicities \citep{Grinsted2004}. 
The continuous wavelet transform of a discrete sequence
$x_n$ is defined as the convolution of $x_n$ (with $n=1,\ldots,N$ and uniform time intervals of size $\delta t$)   with a scaled and translated version of a wavelet function $\psi$:
\begin{equation}
    W^X(s,n)=\sum_{n^{\prime}=0 }^{N-1} x_{n^{\prime}}  \psi^{(*)}\left[\frac{(n^{\prime}-n)\delta t}{s}\right]
\end{equation}
where (*) indicates the complex conjugate and $s$ is the wavelet scale. To be ``admissible'' as a wavelet,
the function $\psi$ must have zero mean and be localized in
both time and frequency space. An example is the Morlet wavelet \citep{TorrenceCompo1998}. 
By varying the wavelet scale $s$ and translating along the localized time index $n$, one can construct a picture showing both the amplitude of any features versus the scale and how this amplitude varies with time.
The wavelet transform $W^X(s,n)$ is complex if the wavelet function $\psi$ is complex. Then, the transform can be divided into the amplitude $|W^X(s,n)|$ and the phase, $\tan^{-1} (\Re(W^X(s,n))/\Im(W^X(s,n)))$. The wavelet power spectrum is defined as $|W^X(s,n)|^2$, and the  cone of influence is the  region of the wavelet spectrum in which edge effects become important  \citep{TorrenceCompo1998}. It is the consequence of dealing with finite length time series, where errors occur at the beginning and end of the wavelet transform \citep{TorrenceCompo1998}.

Wavelet analysis is also a useful tool for analyzing two time series together, to examine whether they are linked in some way. 

From both  CWTs,  the cross wavelet transform  (XWT) and the wavelet coherence (WTC)  were constructed for examining relationships in time--frequency space between two time series  \citep{TorrenceCompo1998}. These tools allow the recognition of common power and relative phase in space time-frequency, respectively, in addition to the evaluation of significant coherence and confidence levels against red (Brownian) noise  \citep{TorrenceCompo1998}. 

Given two time series $x_{t}$ and $y_{t}$ (with $t=1,\ldots,N$ and uniform time intervals of size $\delta t$), the so-called  cross wavelet transform of their respective (continuous) wavelet transforms $W^{X}$and $W^{Y}$ is a 2-D representation of complex numbers given by
\begin{equation}
  W^{XY}=W^{X}(W^{Y})^{\ast} , 
\end{equation}
and the cross wavelet power XWT is defined as $|W^{XY}|$. As the XWT is calculated by multiplying the CWT of the first time series by the complex conjugate of the CWT of the second time series, its absolute value  will be high in the time-frequency areas where both CWTs display high values, thus allowing to identify common temporal patterns in the two data sets. Likewise, the phase of the complex number $W^{XY}$ gives information about the phase relation between $X$ and $Y$ in the time-frequency space.  It is represented by the angle formed by the arrow measured counterclockwise and indicates the time delay between the two time series. For example, it will be 0\textdegree $\ $ (the arrow points to the right) when the two time series are in phase, while it will be around 180\textdegree $\ $ if they are in anti-phase (i.e., one reaches its maximum value when the other reaches the minimum and vice versa). For intermediate phase values the time lag between both series can be calculated using the expression:
\begin{equation}
 \Delta t= \Delta_\phi \times \frac{T}{2\pi}
 \label{eq:time-lag} 
\end{equation}
with $T$ being the period and $\Delta_\phi$ being the gap between the two time series expressed in radians.

Another useful measure is the wavelet coherence, which is given by \cite{TorrenceCompo1998}, 

\begin{equation}
    R^{2}(s,n)=\frac{|S(s^{-1}W^{XY}(s,n))|^{2}}{ S(s^{-1}|W^{X}(s,n)|^{2})S(s^{-1}|W^{Y}(s,n)|^{2})}
    \end{equation}
and is calculated by the normalized cross-correlation between $X$ and $Y$  time series, including a smoothing operator in both domains (time and frequency).

In other words, WTC is the result of the normalization of a smoothed version of the XWT. Its absolute value will be high (i.e., close to 1) in those areas of the temporal frequency plane where the time-frequency pattern is locally similar (hence, coherent) in the two CWTs. 
The interpretation of the phase information is the same as for the XWT, as they are calculated in the same way, with the exception of the smoothing operator used in the WTC. This smoothing operation required for the WTC degrades the resolution of the result in the time-frequency domain.
It can be concluded that XWT and WTC provide  powerful tools for dealing with nonstationary time series. From two CWTs, while XWT reveals high common power, WTC finds locally phase correlated behavior.  As it was mentioned above, WTC is slightly less localized in time frequency space than the XWT. Its significance level is determined using Monte Carlo methods \citep{TorrenceCompo1998}.
A software package was developed by A. Grinsted et al. to perform XWT and WTC \citep{Grinsted2004}.

The application of the continuous wavelet transform (CWT), the cross wavelet transform (XWT) and the wavelet coherence (WTC), both to various synthetic cases and to a real case, shows the usefulness of the proposed methodology for the recognition of seasonal temporal patterns of shocks/instabilities and their relationship (time lags) with events that likely trigger them. 
This methodology has been previously applied in different scientific domains. 
For example, in \citep{Li2016} these tools were used successfully to interpret landslide displacement time series derived from 
Synthetic aperture radar interferometry (InSAR), where localized intermittent periodicities were identified. Another case is \citep{Muchebve18}, which examines  the power of wavelet analysis in nonlinear time series such as salinity intrusion processes. This is important because there  is a lake system on the sea coast of Japan and  seawater frequently intrudes into these lakes. Another application of these wavelet-based tools is presented in \cite{KPVL2021}. In this article, the COVID-19 time series for different countries were analyzed.  Wavelet coherence between the time series of daily new cases and daily deaths in a given country was calculated and was used to quantify changes in the dynamics of the time series, such as the effects of health policies and vaccination campaigns.

In this paper, we apply these  wavelet-based tools to the study of commodity time series in relation with a proxy of economic uncertainty. Considering the aforementioned difference between XWT and WTC, we pay attention to those regions where the graphics of XWT and WTC exhibit high power. In other words, we will work, simultaneously, with both graphics. 

\begin{figure}[!htbp]
\centering
\subfloat[]{\includegraphics[width = 1\textwidth]{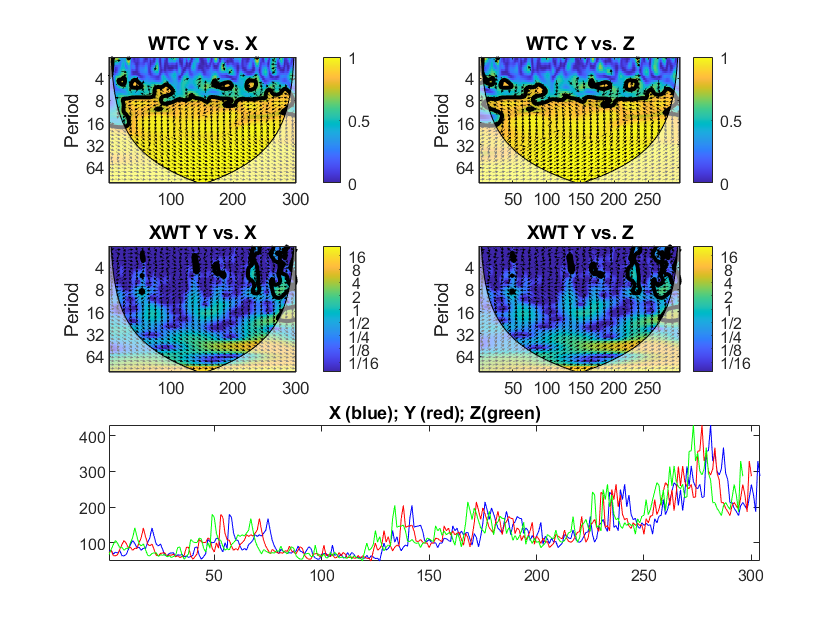}}
\caption{Interpretation of the lead/lag relationship according to arrows' orientation in graphics of  XWT and WTC, for any time series $X$, and its shifted realizations $Y$ and $Z$}
\label{fig:arrow}
\end{figure}

The interpretation of the lead/lag relationship of two time series, related to the orientation of the arrows is as follows. Let's consider a time series $Y$. We construct series $X$ and $Z$, which are shifted realizations of $Y$, such that $X_t=Y_{t-4}$ and $Z_t=Y_{t+4}$. Thus, $Y$ lags $X$ and $Y$ leads $Z$. We construct WTC and XWT graphs of $Y$ vs. $X$ and  $Y$ vs. $Z$, as shown in Figure \ref{fig:arrow}. It can be observed that when the arrows point upwards (1st and 2nd quadrants) the first series leads the second and vice versa.

\subsection{Application: analysis Crude oil}

In this section we perform a detailed analysis of the crude oil and GEPU time series, encompassing the information obtained by the continuous wavelet power spectrum (given by the CWT), the cross wavelet transform, and squared wavelet coherence.

\begin{figure}[!htbp]
\centering
\subfloat[]{\includegraphics[width = 0.5\textwidth]{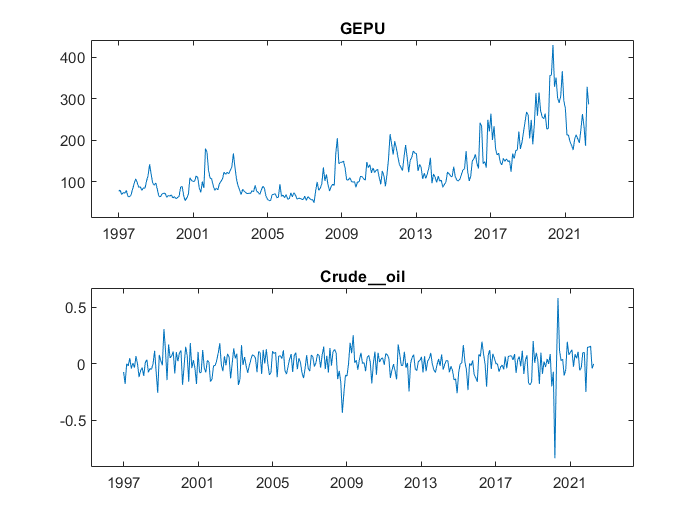}}
\subfloat[]{\includegraphics[width = 0.45\textwidth]{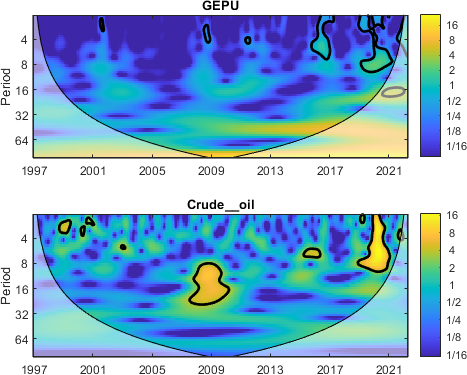}}\\
\caption{Crude oil returns and GEPU time series (a) and their continuous wavelet power spectrum.}
\label{fig:wavelet}
\end{figure}

Figure \ref{fig:wavelet} shows GEPU and crude oil time series (on the left) and the power spectrum (on the right) provided by the respective wavelet transforms. We observe in yellow and light green/blue, the areas where wavelet power is significant. Regarding crude oil, we observe a peak in the 8-16 period around 2009, and another one in the 1-8 period in 2020. Whereas, for GEPU it can be observed a less pronounced (but significant) peak in the 1-4 period in 2017 and 2020. 

\begin{figure}[!htbp]
\centering
\subfloat[]{\includegraphics[width = 0.45\textwidth]{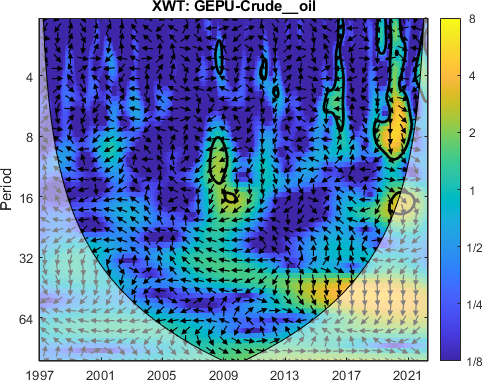}\label{fig:xwt_crude}} 
\subfloat[]{\includegraphics[width = 0.45\textwidth]{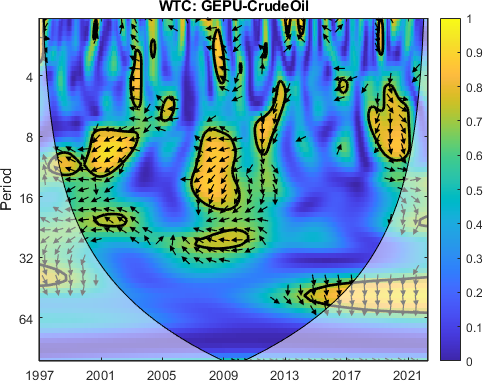}\label{fig:wtc_crude}}
\caption{Cross wavelet transform (left) and wavelet coherence (right) of the standardized crude oil and GEPU time series.}
\label{fig:CrudeOilcross}
\end{figure}

Figure \ref{fig:xwt_crude} shows the cross wavelet transform (XWT) and depicts the areas with high common power between GEPU and crude oil. Such areas appear in warm colors (especially yellow) and are contoured by thick black lines, conforming a closed region. This figure reveals that crude oil and GEPU share significant power for the 8-14 period in 2008-2009, for the 2-10 period in 2020, and (to a lesser extent) for the 2-9 period in 2017. The arrows' angles indicate the phase delay. These phase angles are not constant across scales, which is a signature of varying time lag in signal propagation. 

Finally, Figure \ref{fig:wtc_crude} shows the WTC, which represents the coherence in the time-frequency space of both signals. There we observe a significant coherence in the 8-16 period around 2008-2009 and in the 4-10 period around 2020. Our analysis encompasses the results of both the XWT and the WTC. In other words, we consider the significant coherent areas, only when XWT is also significant. Therefore, we will consider (for these two time series) only the coherence and delay in 2008-2009 and in 2020. 

If we focus our attention in the years 2008-2009, considering the overlapping areas of XWT and WTC and the arrows' orientation, the phase delay between GEPU and crude oil (following Equation  \ref{eq:time-lag}) is between 5 and 6 months for high-frequency analysis and between 6 and 10 months for low-frequency analysis. Additionally, if we now consider the year 2020, the analysis reveals a phase delay between both time series between 5 to 6 months at high frequency and between 6 to 8 months at low frequency. In all cases, GEPU lags crude oil for the mentioned delay periods.

\section{Data and results \label{sec:DataResults}}
This paper uses monthly data of 16 commodities and the GEPU index from January 1997 until April 2022, with a total of 304 observations. Selected commodities range from ferrous, non-ferrous, and precious metals, to food and energy commodities:  copper, aluminium, nickel, silver, gold, palladium, platinum, cotton, wheat, cocoa, coffee, raw sugar, corn, soyabeans, crude oil, and heating oil. 
The GEPU index (global) is the Gross Domestic Product weighted average of the economic and policy uncertainty indices of 21 countries. It is based upon the relative frequency of newspaper articles of each country, targeting terms that proxy economic, policy or uncertainty news \citep{Davis2016}. 
Commodities' data were downloaded from Eikon and GEPU index from its website.

The period under study is crossed by different crises and stressful situations. These events are assumed to be captured by the GEPU index. Due to their different nature, intensity, and particular characteristics, we could expect commodities to react differently to uncertainty. That is why our analysis has proposed to distinguish and cluster different behaviors, classifying them according to the most important crises identified in the period examined. We select three important crises, each of them of different nature. The first one is the 2001 dot-com crisis, which corresponds to a  technology-inspired boom in the late 1990s. The crisis is reflected in the sustained downturn of the NASDAQ composite index from its peak in March 2000 and the decreases were later accentuated by the events of September 11. As such, it can be considered a crisis in a specific sector of the economy, aggravated by a political event.  The second one is the 2008 Global Financial Crisis, which is the first worldwide crisis of the 21st. century and affected many countries. Finally, the third one is the Covid-19 pandemic. Unlike previous events, this event could be considered an unforeseeable, one-in-a-life event. Although  a health emergency, it produced important economic effects, such as supply chain disruptions and a sudden stop in many economic activities across the world. Then, our analysis precisely sheds light on the different linkages between GEPU and commodities during such events.

At a general level, and looking only at the WTC crude-oil and heating oil present very significant and similar regions during the three crises (considering the amount colored in yellow). In the case of corn, raw sugar, and wheat, what the WTC shows in the three crises is also similar behavior, but much less significant. In the case of cotton and platinum, the WTC is similar for the 3 crises with medium significance. In the rest of the commodities, the WTC is much more significant in the 2009 crisis than in the other 2 crises.

\begin{figure}[!htbp]
\centering
\subfloat[]{\includegraphics[width = 0.45\textwidth]{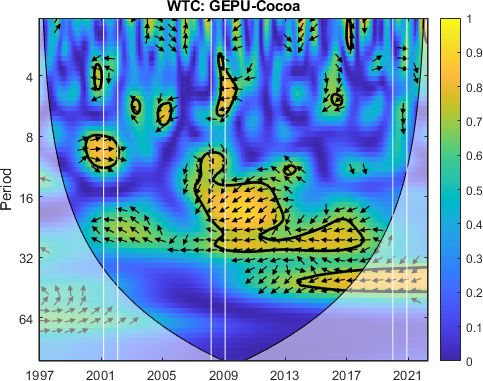}}
\subfloat[]{\includegraphics[width = 0.45\textwidth]{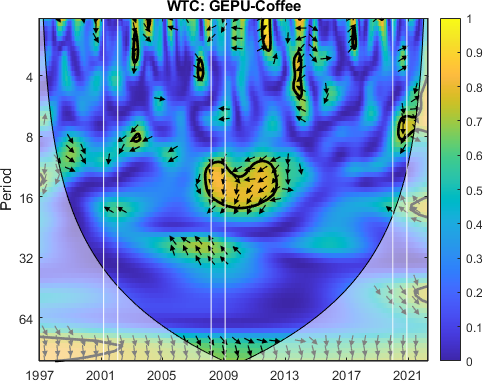}}\\
\subfloat[]{\includegraphics[width = 0.45\textwidth]{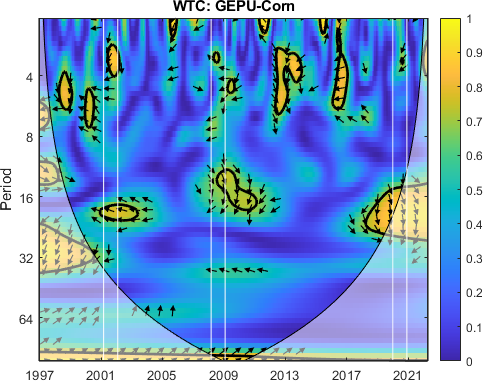}}
\subfloat[]{\includegraphics[width = 0.45\textwidth]{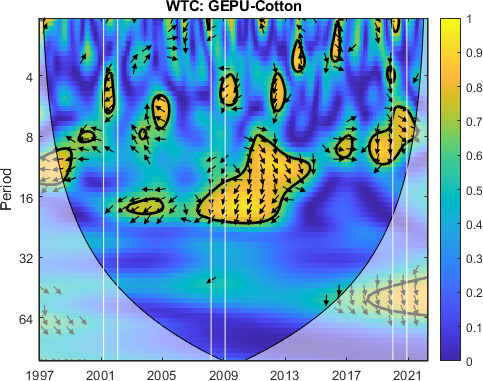}}\\
\subfloat[]{\includegraphics[width = 0.45\textwidth]{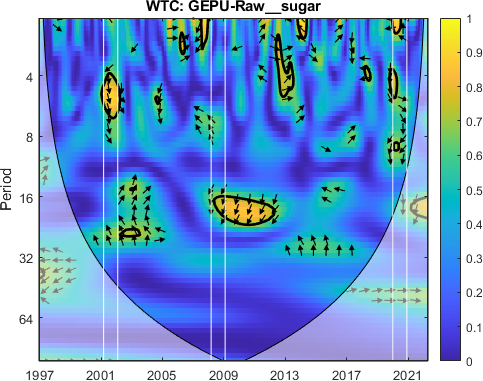}}
\subfloat[]{\includegraphics[width = 0.45\textwidth]{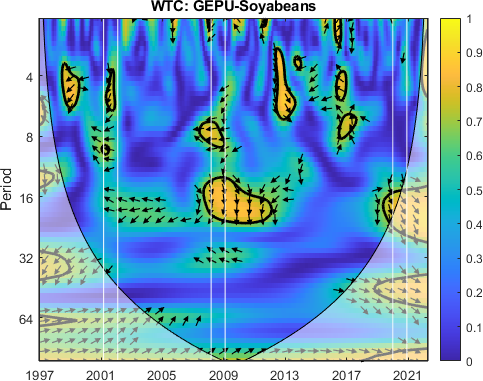}}\\
\subfloat[]{\includegraphics[width = 0.45\textwidth]{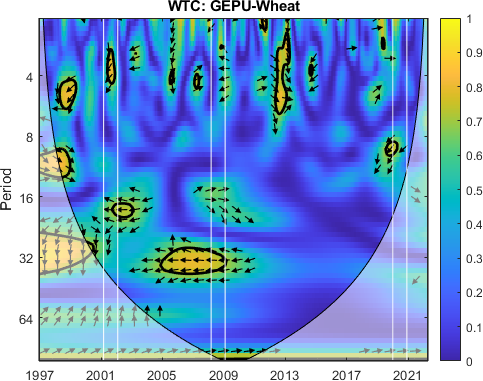}}
\caption{Wavelet Coherence (WTC) between GEPU and agricultural commoditties.}
\label{fig:agriculturalWTC}
\end{figure}

\begin{figure}[!htbp]
\centering
\subfloat[]{\includegraphics[width = 0.45\textwidth]{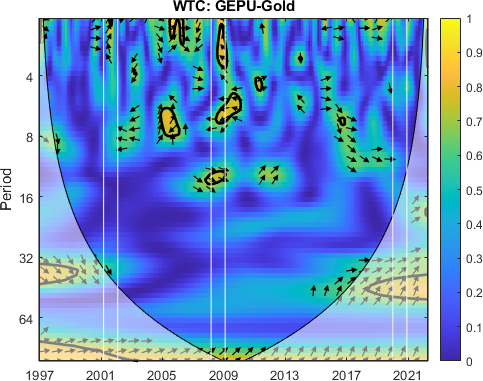}}
\subfloat[]{\includegraphics[width = 0.45\textwidth]{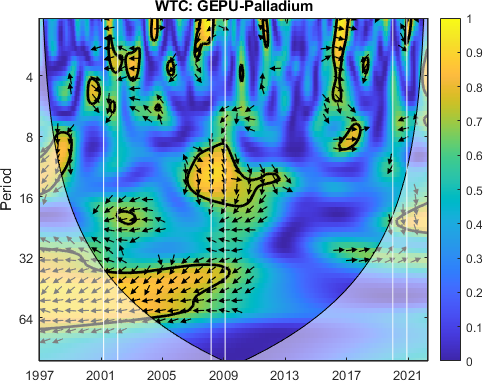}}\\
\subfloat[]{\includegraphics[width = 0.45\textwidth]{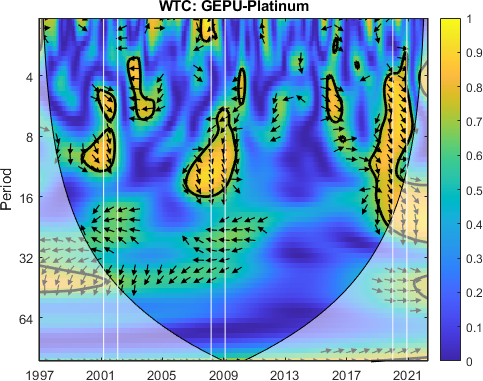}}
\subfloat[]{\includegraphics[width = 0.45\textwidth]{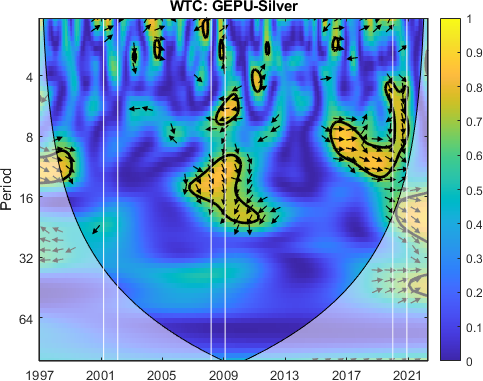}}
\caption{Wavelet Coherence (WTC) between GEPU and precious metals.}
\label{fig:preciousWTC}
\end{figure}

\begin{figure}[!htbp]
\centering
\subfloat[]{\includegraphics[width = 0.45\textwidth]{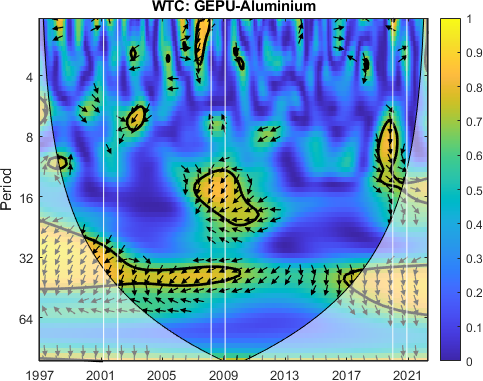}}
\subfloat[]{\includegraphics[width = 0.45\textwidth]{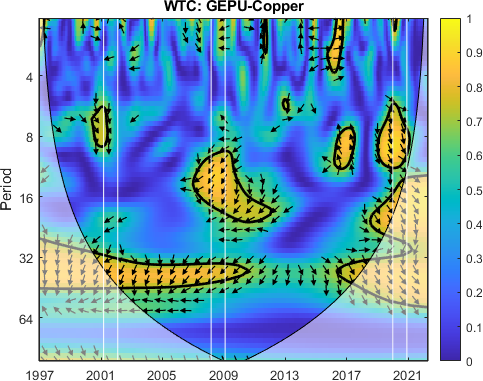}}\\
\subfloat[]{\includegraphics[width = 0.45\textwidth]{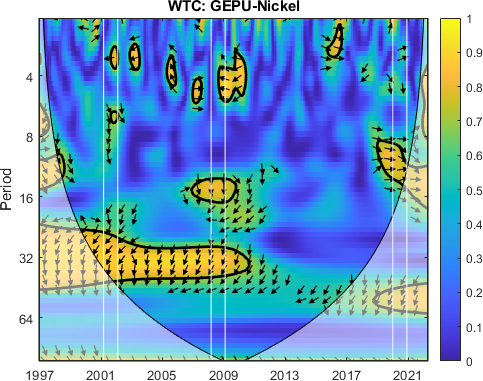}}
\caption{Wavelet Coherence (WTC) between GEPU and non-precious metals.}
\label{fig:nonpreciousWTC}
\end{figure}

\begin{figure}[!htbp]
\centering
\subfloat[]{\includegraphics[width = 0.45\textwidth]{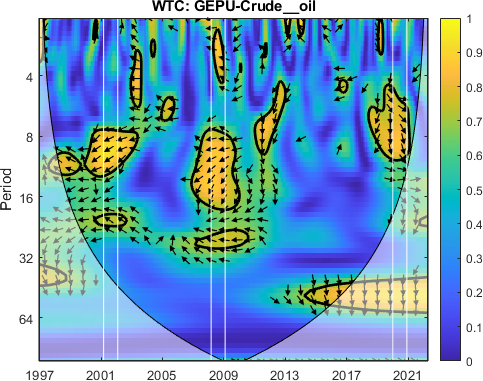}}
\subfloat[]{\includegraphics[width = 0.45\textwidth]{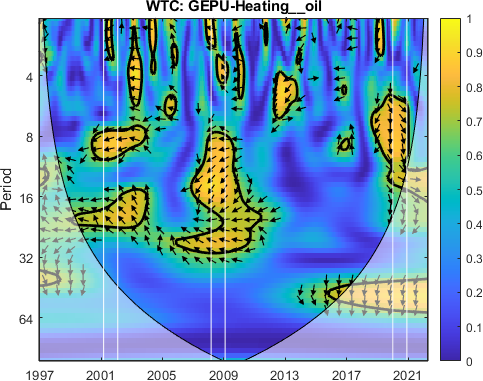}}
\caption{Wavelet Coherence (WTC) between GEPU and energy.}
\label{fig:energyWTC}
\end{figure}

\subsection{The Dot-com crisis}

In the year 2001, two events converge. On one side the burst of the so-called dot-com bubble and the 9/11 attack. In addition to a profound downturn in the NASDAQ Composite Index, it produced a reduction in the venture capital available for financing innovative firms, and a change in the mentality of CEOs and investors, regarding tech firms. Moreover, the  September 11 attacks contributed further to an increase in global and political uncertainty.

Table \ref{tab:summaryHigh} and Table \ref{tab:summaryLow} summarize the lead-lag relationships and the delay between GEPU and each one of the commodities in the three selected crises. We consider a significant relationship only in  those regions where the graphics of XWT and WTC exhibit high power. In addition, as explained in Section \ref{sec:methods}, the delay is measured by the angle of the arrows in WTC graphics.

We can observe that cotton, sugar, palladium, platinum, and nickel exhibit coherence with GEPU at high frequency ($T=2-7$) with delays between 1 to 5 months. Additionally, cocoa and heating oil have coherence with GEPU at lower frequencies ($T=8-11$) with immediate delays of up to 6 months. For the remaining commodities, there are no intersection areas between WTC and XWT.
This period is associated with short-term comovements between GEPU and some commodities, but only at higher frequencies (less than 6 months), whereas during other crises the comovements expanded also to lower frequencies.

\begin{table}
\caption{Lead-lag relationship between a given commodity and GEPU, considered at high frequency (less than 6 months). Colours indicate the sense of the relationship: red means that GEPU lags the commodity, whereas blue signals that GEPU leads the commodity.}
\resizebox{\textwidth}{!}{
\begin{tabular}{ l  w{l}{3.5cm}  w{l}{3.5cm}  w{l}{3.5cm} } \hline
Commodities &	  2001   &		2008		&	 2020			\\ \hline
Cotton	   &	\textcolor{red}{2.4-3.8}      &	\textcolor{blue}{2.5-3.7}   &	\textcolor{red}{0.8-1.1}, \textcolor{blue}{2.8-3.2}, \textcolor{blue}{4.6-6}	\\ 
Sugar	     &	\textcolor{blue}{3.1-4.7}	   &\cellcolor{gray!25}     & \textcolor{red}{0.9-1.2}, \textcolor{blue}{2.7-3.9} 	 \\ 
Coffee	   &	\cellcolor{gray!25}&	\cellcolor{gray!25}	&	\textcolor{blue}{5.3-6}		\\ 
Cocoa	     &	\textcolor{blue}{0.1-0.7}	& \textcolor{blue}{2-3.6}	& \cellcolor{gray!25}	\\ 
Wheat	     &\cellcolor{gray!25}				&\cellcolor{gray!25}		&	\textcolor{blue}{5.7-6}	 \\ 
Soyabeans	 &	\cellcolor{gray!25}		&	\cellcolor{gray!25}	&\cellcolor{gray!25}\\ 
Corn	     &	\cellcolor{gray!25}     &	\textcolor{red}{1.1-1.3}	&\cellcolor{gray!25}\\ \hline
Heating Oil&  \textcolor{blue}{5.1-6}	&	\textcolor{red}{0.8-1.2}, \textcolor{blue}{5.1-6}	& 	\textcolor{blue}{4.6-6}	  	\\ 
Crude Oil	 &\cellcolor{gray!25}	&	\textcolor{blue}{4.6-6}	&	 \textcolor{blue}{5-6}	  \\ \hline
Palladium	 &	\textcolor{blue}{1.3-1.9}, \textcolor{blue}{3.6-4} 	& \textcolor{blue}{1.6-1.7}  	&\cellcolor{gray!25}\\ 
Platinum	 &	\textcolor{blue}{3.9-4.8}	&	\textcolor{blue}{3.1-5.7}, \textcolor{blue}{5.9-6 }	&	\textcolor{blue}{3.8-4.8}, \textcolor{blue}{4.3-6} \\ 
Silver	   &\cellcolor{gray!25}	&	 \textcolor{blue}{3.3-4.1}	&	\textcolor{blue}{3.7-6}\\ 

Nickel	   & \textcolor{blue}{1.6-2}	&	\textcolor{blue}{1.6-2.5}  &\cellcolor{gray!25}				\\ 
Copper	   &	\cellcolor{gray!25}	&	\cellcolor{gray!25}	&	 \textcolor{blue}{5.7-6}	\\ 
Aluminium  &\cellcolor{gray!25}	&\cellcolor{gray!25}	&	\textcolor{blue}{5.3-6}	 \\ 
Gold	     &\cellcolor{gray!25}		&	\textcolor{red}{0.6-1.8}, \textcolor{blue}{2.9-4.2}		&	\cellcolor{gray!25}	\\ \hline
\end{tabular}
}
\label{tab:summaryHigh}
\end{table}

\begin{table}
\caption{Lead-lag relationship between a given commodity and GEPU, considering at low frequency (greater than 6 months). Colours indicate the sense of the relationship: red means that GEPU lags the commodity, whereas blue signals that GEPU leads the commodity. }

\begin{tabular}{l  w{l}{3cm}  w{l}{3cm}  w{l}{3cm} } \hline
Commodities &	  2001   &	 2008		&	2020			\\ \hline
Cotton	   &\cellcolor{gray!25}       	&\cellcolor{gray!25}       &	  \textcolor{blue}{6-6.9}	       \\ 
Sugar	     &	  \cellcolor{gray!25}    & \textcolor{blue}{11.2-11.8}	        &	\textcolor{blue}{6.1-6.4}	      \\ 
Coffee	   &\cellcolor{gray!25}	&	\cellcolor{gray!25}	&	\cellcolor{gray!25}	\\ 
Cocoa	     &	\cellcolor{gray!25}	&	\cellcolor{gray!25}		& \cellcolor{gray!25}	\\ 
Wheat	     &	\cellcolor{gray!25}		&	\cellcolor{gray!25}		&	\textcolor{blue}{6-6.2} \\ 
Soyabeans	 &	\cellcolor{gray!25}	& \textcolor{blue}{11.2-12.6} &	\cellcolor{gray!25} \\ 
Corn	     &	\cellcolor{gray!25}	   & \cellcolor{gray!25}	&	\textcolor{blue}{11-14.5} \\ \hline
Heating Oil & \textcolor{blue}{6-6.1}	&	 \textcolor{blue}{6-7.9}	& 	\textcolor{blue}{6-7.6}	    \\ 
Crude Oil	 & \cellcolor{gray!25}	&	\textcolor{blue}{6-9.7}	&	 \textcolor{blue}{6-8.1}  \\ \hline
Palladium	 &\cellcolor{gray!25}	 &	 \textcolor{blue}{8.1-9.6}	&\cellcolor{gray!25}		\\ 
Platinum  &	\cellcolor{gray!25}  &	 \textcolor{blue}{6-12.5 }	&	 \textcolor{blue}{6-7.2},  \textcolor{blue}{12.9-14.8}  \\ 
Silver	   &\cellcolor{gray!25}	       &   \cellcolor{gray!25}   	  &	\textcolor{blue}{6-8}	\\ 
Nickel	   & \cellcolor{gray!25}  &\cellcolor{gray!25} 	&	\textcolor{blue}{8.5-9.6}				\\ 
Copper	   &\cellcolor{gray!25}			&	\textcolor{blue}{6.3-9.8}		&	 \textcolor{blue}{6-7.2}	\\ 
Aluminium  &\cellcolor{gray!25}		&	\textcolor{blue}{8.1-11.2}	&	\textcolor{blue}{6-7.5}\\ 
Gold	     &	\cellcolor{gray!25}	&	\cellcolor{gray!25}  &\cellcolor{gray!25}\\ \hline
\end{tabular} \\
\label{tab:summaryLow}
\end{table}

\subsection{The 2008 financial crisis}
The 2008 Global Financial Crisis (GFC) is undoubtedly the most important distress event in the 21st century. From the comparison of results shown in Figures \ref{fig:agriculturalWTC} to \ref{fig:energyWTC}  and in Tables \ref{tab:summaryHigh} and \ref{tab:summaryLow}, the wavelet coherence of GEPU with respect to the commodities can be observed. Cotton, cocoa, corn, silver, nickel, and gold show coherence at high frequencies ($T=2-7$), with a lag of about 4 months. On the other hand, sugar, soyabeans, copper, and aluminum show common behavior at low frequencies ( $T=10-19$), with a lag of 6 to 13 months. Crude oil shows a coherence with GEPU at low frequencies ($T=8-14$), with a lag between 5 to 9 months, while heating oil, palladium, and platinum show coherence at both high ($T=2-8$), and low frequencies ($T=8-16$), with lags ranging from 5 to 12 months.

\subsection{Covid pandemic}

Encompassing the visual analysis of Figures \ref{fig:agriculturalWTC} to \ref{fig:energyWTC} and the summarized information in Tables \ref{tab:summaryHigh} and \ref{tab:summaryLow}, we detect that cotton, sugar, heating oil y silver show coherent behavior with GEPU at high frequencies ($T=2-7$) with delays less than 5 months. Contrarily, cotton, sugar, coffee, wheat, heating oil, crude oil, palladium, silver, copper, nickel, and aluminium display coherent behavior at lower frequencies ($T=7-10$) with delays between 5 to 8 months, except for nickel, whose delay is around 9 months. Moreover, corn displays coherence at even lower frequencies ($T=14-19$) with delays between 11 to 15 months. Finally platinum features both high ($T=3-10$) and low ($T=16-19$) frequencies coherence, with delays of 3 to 7 and 13 to 15 months, respectively. 

\section{Discussion and conclusions \label{sec:conclusions}}
\begin{figure}[!htbp]
\centering
\includegraphics[width=\textwidth]{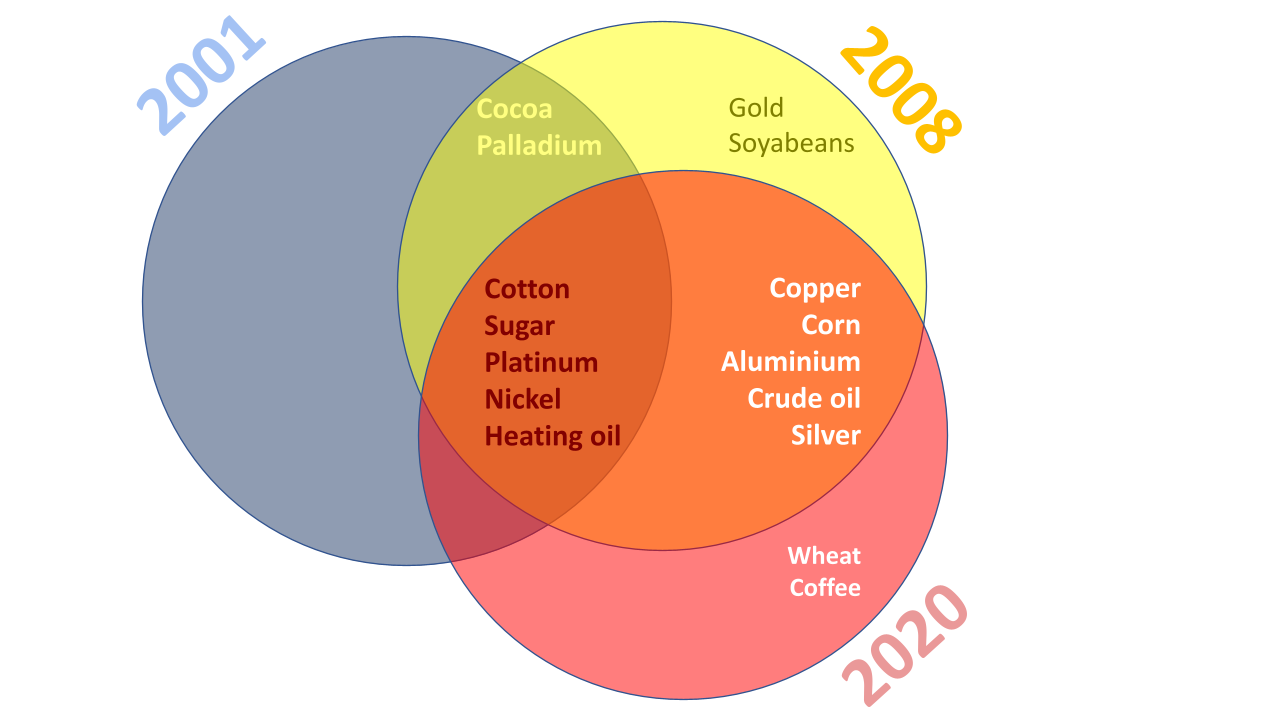}
\caption{Venn diagram of the significant coherence between GEPU and each of the commodities, during the three crises considered.}
\label{fig:Diag}
\end{figure}

This paper discussed the time-frequency relationship between GEPU and different commodities. Our analysis reveals that not all crises affect commodities in the same way. In fact, the Global Financial Crisis was the event that is most closely related to coherent behavior between GEPU and commodities. 
To summarize our findings, we show a Venn diagram in Figure \ref{fig:Diag}, where each set represents one of the crises considered, and the names of the commodities are those that exhibit coherence with GEPU during the respective crisis. 
We observe that the Global Financial Crisis of 2008 and the Covid-19 crisis were the most significant events in our sample, where 16 commodities were affected by  economic uncertainty, whereas the 2001 crisis affected only the behavior of 7 commodities. Even more, it is important to point out that the composition of the commodities affected by the different crises was not the same. It must be highlighted that the GFC was the single crisis where gold dynamics reflected some commonality with GEPU, albeit during a very short time.

As its shown in Figure \ref{fig:Diag} during the 2001 crisis GEPU exhibits coherent behavior with some agricultural commodities and metals. However, neither important agricultural products such as wheat and soyabeans, nor precious metals such as gold and silver were affected.

Finally, gold is mainly detached from the evolution of the general economy. In fact, the only period where coherence was detected was during the Global Financial Crisis, but only for a short period, as can be seen in Table \ref{tab:summaryHigh}. Also soyabeans were only affected by the GFC, but for a longer period (almost 13 months). The only commodities that do not display comovements with GEPU are wheat and coffee. Unlike previous crises, the Covid-19 emergency did not produce a coherence between GEPU and gold or palladium, soyabeans, and cocoa. This could be due to the quick reaction of different governments, providing financial aid and subsidies to individuals and firms, to alleviate the effects of lockdowns and possible economic downturns.

\bibliographystyle{abbrvnat}
\bibliography{commoditybib}

\begin{thebibliography}{15}
\providecommand{\natexlab}[1]{#1}
\providecommand{\url}[1]{\texttt{#1}}
\expandafter\ifx\csname urlstyle\endcsname\relax
  \providecommand{\doi}[1]{doi: #1}\else
  \providecommand{\doi}{doi: \begingroup \urlstyle{rm}\Url}\fi

\bibitem[Adams and Gl{\"{u}}ck(2015)]{Adams2015}
Z.~Adams and T.~Gl{\"{u}}ck.
\newblock {Financialization in commodity markets: A passing trend or the new
  normal?}
\newblock \emph{Journal of Banking \& Finance}, 60:\penalty0 93--111, nov 2015.
\newblock ISSN 03784266.
\newblock \doi{10.1016/j.jbankfin.2015.07.008}.
\newblock URL
  \url{https://linkinghub.elsevier.com/retrieve/pii/S0378426615002022}.

\bibitem[Batten et~al.(2010)Batten, Ciner, and Lucey]{Batten2010}
J.~A. Batten, C.~Ciner, and B.~M. Lucey.
\newblock {The macroeconomic determinants of volatility in precious metals
  markets}.
\newblock \emph{Resources Policy}, 35\penalty0 (2):\penalty0 65--71, jun 2010.
\newblock ISSN 03014207.
\newblock \doi{10.1016/j.resourpol.2009.12.002}.
\newblock URL
  \url{https://linkinghub.elsevier.com/retrieve/pii/S0301420709000543}.

\bibitem[Chen(2015)]{Chen2015}
P.~Chen.
\newblock {Global oil prices, macroeconomic fundamentals and China's commodity
  sector comovements}.
\newblock \emph{Energy Policy}, 87:\penalty0 284--294, dec 2015.
\newblock ISSN 03014215.
\newblock \doi{10.1016/j.enpol.2015.09.024}.
\newblock URL
  \url{https://linkinghub.elsevier.com/retrieve/pii/S0301421515301105}.

\bibitem[Davis(2016)]{Davis2016}
S.~J. Davis.
\newblock {An Index of Global Economic Policy Uncertainty}, 2016.
\newblock URL \url{http://www.nber.org/papers/w22740}.

\bibitem[Erb and Harvey(2006)]{Erb2006}
C.~B. Erb and C.~R. Harvey.
\newblock {The Strategic and Tactical Value of Commodity Futures}.
\newblock \emph{Financial Analysts Journal}, 62\penalty0 (2):\penalty0 69--97,
  mar 2006.
\newblock ISSN 0015-198X.
\newblock \doi{10.2469/faj.v62.n2.4084}.
\newblock URL
  \url{https://www.tandfonline.com/doi/full/10.2469/faj.v62.n2.4084}.

\bibitem[Fernandes et~al.(2022)Fernandes, de~Araujo, Silva, and
  Tabak]{FernandesAraujoSilvaTabak2022}
L.~H. Fernandes, F.~H. de~Araujo, J.~W. Silva, and B.~M. Tabak.
\newblock {Booms in commodities price: Assessing disorder and similarity over
  economic cycles}.
\newblock \emph{Resources Policy}, 79:\penalty0 103020, dec 2022.
\newblock ISSN 03014207.
\newblock \doi{10.1016/j.resourpol.2022.103020}.
\newblock URL
  \url{https://linkinghub.elsevier.com/retrieve/pii/S0301420722004639}.

\bibitem[Grinsted et~al.(2004)Grinsted, Moore, and Jevrejeva]{Grinsted2004}
A.~Grinsted, J.~C. Moore, and S.~Jevrejeva.
\newblock {Application of the cross wavelet transform and wavelet coherence to
  geophysical time series}.
\newblock \emph{Nonlinear Processes in Geophysics}, 11\penalty0 (5/6):\penalty0
  561--566, nov 2004.
\newblock ISSN 1607-7946.
\newblock \doi{10.5194/npg-11-561-2004}.
\newblock URL \url{https://npg.copernicus.org/articles/11/561/2004/}.

\bibitem[{International Copper Association (ICA)}()]{CopperChile}
{International Copper Association (ICA)}.
\newblock The socioeconomic impact of copper mining in {C}hile.
\newblock URL
  \url{https://copperalliance.org/resource/the-socioeconomic-impact-of-copper-mining-in-chile}.
\newblock Available online at:
  \url{https://copperalliance.org/resource/the-socioeconomic-impact-of-copper-mining-in-chile},
  accessed: 24/11/2022.

\bibitem[Irwin and Sanders(2010)]{IrwinSanders}
S.~H. Irwin and D.~R. Sanders.
\newblock {The Impact of Index and Swap Funds on Commodity Futures Markets:
  Preliminary Results}.
\newblock OECD Food, Agriculture and Fisheries Papers~27, OECD Publishing, June
  2010.
\newblock URL \url{https://ideas.repec.org/p/oec/agraaa/27-en.html}.

\bibitem[Kowalski et~al.(2023)Kowalski, Portesi, Vampa, Losada, and
  Holik]{KPVL2021}
A.~Kowalski, M.~Portesi, V.~Vampa, M.~Losada, and F.~Holik.
\newblock Information quantifiers and unpredictability in the covid-19
  time-series data.
\newblock \emph{Revista de Matemática: Teoría y Aplicaciones - cimpa – ucr
  issn: 1409-2433 (Print), 2215-3373 (Online)}, forthcoming, 10 2023.

\bibitem[Muchebve et~al.(2018)Muchebve, Nakamura, and Kamiya]{Muchebve18}
E.~Muchebve, Y.~Nakamura, and H.~Kamiya.
\newblock Use of wavelet techniques in the study of seawater flux dynamics in
  coastal lakes.
\newblock In S.~Radhakrishnan, editor, \emph{Wavelet Theory and Its
  Applications}, chapter~11. IntechOpen, Rijeka, 2018.
\newblock \doi{10.5772/intechopen.75177}.
\newblock URL \url{https://doi.org/10.5772/intechopen.75177}.

\bibitem[Nguyen et~al.(2021)Nguyen, Naeem, Balli, Balli, and Vo]{Nguyen2021}
T.~T.~H. Nguyen, M.~A. Naeem, F.~Balli, H.~O. Balli, and X.~V. Vo.
\newblock {Time-frequency comovement among green bonds, stocks, commodities,
  clean energy, and conventional bonds}.
\newblock \emph{Finance Research Letters}, 40:\penalty0 101739, may 2021.
\newblock ISSN 15446123.
\newblock \doi{10.1016/j.frl.2020.101739}.
\newblock URL
  \url{https://linkinghub.elsevier.com/retrieve/pii/S1544612320304207}.

\bibitem[Stoll and Whaley(2010)]{Stoll2010}
H.~R. Stoll and R.~E. Whaley.
\newblock {Commodity Index Investing and Commodity Futures Prices}.
\newblock \emph{Journal of Applied Finance}, 20\penalty0 (1):\penalty0 7--46,
  2010.
\newblock ISSN 15346668.

\bibitem[Tomás et~al.(2016)Tomás, Li, Lopez-Sanchez, Liu, and
  Singleton]{Li2016}
R.~Tomás, Z.~Li, J.~Lopez-Sanchez, P.~Liu, and A.~Singleton.
\newblock Using wavelet tools to analyse seasonal variations from insar
  time-series data: a case study of the huangtupo landslide.
\newblock \emph{Landslides}, 13:\penalty0 437--450, 06 2016.
\newblock \doi{10.1007/s10346-015-0589-y}.

\bibitem[Torrence and Compo(1998)]{TorrenceCompo1998}
C.~Torrence and G.~P. Compo.
\newblock {A Practical Guide to Wavelet Analysis}.
\newblock \emph{Bulletin of the American Meteorological Society}, 79\penalty0
  (1):\penalty0 61--78, jan 1998.
\newblock ISSN 0003-0007.
\newblock \doi{10.1175/1520-0477(1998)079<0061:APGTWA>2.0.CO;2}.
\newblock URL
  \url{http://journals.ametsoc.org/doi/10.1175/1520-0477(1998)079%3C0061:APGTWA%3E2.0.CO;2}.

\end{thebibliography}

\end{document}